\documentclass[journal]{IEEEtran}

\ifCLASSINFOpdf
  \usepackage[pdftex]{graphicx}
  \DeclareGraphicsExtensions{.pdf,.jpeg,.png}
\else
  \usepackage[dvips]{graphicx}
\fi
\usepackage{subfigure}
\usepackage{epstopdf}
\usepackage[cmex10]{amsmath}
\usepackage{amssymb}
\usepackage{wasysym}
\usepackage{algorithm}
\usepackage{algorithmic}
\usepackage{array}
\usepackage{cite}
\usepackage{color}
\usepackage{url}
\usepackage{verbatim}
\usepackage{booktabs}
\usepackage{epsfig,latexsym}
\usepackage{flushend}
\usepackage{subeqnarray}
\usepackage{cases}
\usepackage{cite}
\usepackage{hyperref}
\usepackage{makecell}
\usepackage{amsmath}
\usepackage{stfloats}

\makeatletter
\renewcommand\normalsize{%
   \abovedisplayskip 2\p@ \@plus2\p@ \@minus5\p@
   \abovedisplayshortskip \z@ \@plus3\p@
   \belowdisplayshortskip 2\p@ \@plus3\p@ \@minus3\p@
   \belowdisplayskip \abovedisplayskip
   \let\@listi\@listI}
\makeatother

\begin{document}
%
\title{Energy Efficient Multicast Precoding for Multiuser Multibeam Satellite Communications}


\author{Chenhao~Qi,~\IEEEmembership{Senior~Member,~IEEE}, Huajian~Chen,~\IEEEmembership{Student~Member,~IEEE} \\ Yansha~Deng,~\IEEEmembership{Member,~IEEE}, and Arumugam Nallanathan,~\IEEEmembership{Fellow,~IEEE}
\thanks{This work was supported in part by National Natural Science Foundation of China under Grant 61871119, by Natural Science Foundation of Jiangsu Province under Grant BK20161428, and by the Fundamental Research Funds for the Central Universities. (\textit{Corresponding author: Chenhao~Qi})}
\thanks{Chenhao~Qi and Huajian~Chen are with the School of Information Science and Engineering, Southeast University, Nanjing 210096, China (Email: qch@seu.edu.cn).}
\thanks{Yansha~Deng is with the Department of Informatics, King’s College London, UK (Email: yansha.deng@kcl.ac.uk).}
\thanks{Arumugam Nallanathan is with the School of Electronic Engineering and Computer Science, Queen Mary University of London, UK (Email: a.nallanathan@qmul.ac.uk).}
}

\markboth{Accepted by IEEE Wireless Communications Letters}
{Shell \MakeLowercase{\textit{et al.}}: Bare Demo of IEEEtran.cls for Journals}
\maketitle

\begin{abstract}
Aiming at maximizing the energy efficiency (EE) of multicast multibeam satellite communications, we consider the precoding design under the total power and quality of service (QoS) constraints. Since the original EE maximization problem is nonconvex, it is sequentially converted into a concave-convex fractional programming problem by introducing some variables and using the first-order Taylor low bound approximation. Based on the Charnes-Cooper transformation, it is further converted into a convex problem. Then an iterative algorithm is presented to design the energy efficient precoding. To find feasible initialization points for the algorithm and ensure its convergence, another convex optimization problem with some nonnegative slack variables and a positive penalty parameter is iteratively solved. In particular, the algorithm is verified by the measured channel data of multibeam satellite communications.
\end{abstract}

\section{Introduction}
The explosive demand on wireless access and service prompts the fast development of wireless communications. To provide seamless wireless signal coverage for highland, mountainous and sea areas, we typically resort to the satellite communications, which are highly matched with the particularity of these areas. Aside of providing wide signal coverage, satellite communications are also well known for the strong capability of emergency communications. Therefore, satellite communications are recognized as important parts for 5G and beyond. Recently, precoding design for multibeam satellite communications is attracting more interests, since the precoding can mitigate the multibeam interference as well as compensating the channel fading and therefore can improve the spectral efficiency (SE)~\cite{Vazquez2015Precoding}. In order to provide larger bandwidth for each user, full frequency reuse among different beams can be used, which is generally adopted in terrestrial wireless communications. However, it may result in the inter-beam interference, which is also known as the co-channel interference in the satellite community. As a consequence, precoding is required to mitigate the inter-beam interference. In \cite{Christopoulos2014Multicast}, a frame-based precoding design scheme under per-antenna power constraints for multicast multibeam satellite systems is proposed. In~\cite{Joroughi2017Generalized}, a two-stage low-complexity precoding design scheme for multibeam multicast satellite systems is proposed, where the first stage minimizes the inter-beam interference and the second stage enhances the intra-beam signal-to-interference-and-noise ratio (SINR). However, both of them consider the precoding design aiming at the SE maximization while neglecting the energy efficiency (EE) aspect. In~\cite{qi2018precoding}, based on zero-forcing and sequential convex approximation (SCA), two precoding design algorithms for EE maximization problem under the total power and the quality of service (QoS) constraints for multibeam satellite communications are presented. However, it only focuses on the unicast scenario while lacking the consideration of the multicast scenario.

In this letter, aiming at maximizing the EE, we consider the precoding design for multicast multibeam satellite communications under the total power and QoS constraints. Since the original EE maximization problem is highly nonconvex, we sequentially convert it into a concave-convex fractional programming problem by introducing some variables and using the first-order Taylor low bound approximation. Based on the Charnes-Cooper transformation, we further convert it into a convex problem. Then we present an iterative algorithm to design the energy efficient precoding. To find feasible initialization points for the algorithm and ensure its convergence, another
convex optimization problem with some nonnegative slack variables and a positive penalty parameter is iteratively solved. In particular, the algorithm is verified by the measured channel data of multibeam satellite communications.


\section{System Model}\label{sec.SystemModel}
We consider a multibeam satellite communication system with a single broadband multibeam satellite, a gateway and several users. The gateway transmits the information to the satellite via the feeder link. Since the gateway is equipped with large-size antennas offering high antenna gain, and is normally fixed and placed in an open area with the line-of-sight (LOS) channel to the satellite, we assume the feeder link to be an ideal channel~\cite{Joroughi2017Generalized}. The array fed reflector on the satellite transforms $M$ feeder signals into $N$ transmit signals, where the coverage of each signal on the earth ground forms a beam and there are totally $N$ beams.


Suppose each beam of the satellite covers $Q$ ground users~\footnote{If the number of users per beam is different, supposing the largest number per beam is $Q_{\rm max}$, we may set $Q = Q_{\rm max}$ so that each beam has some virtual users with zero channel coefficients.}. Then the satellite serves totally $K \triangleq QN$ users. We denote the signal vector received by the $K$ users as $\boldsymbol{y} \in \mathbb{C}^{K}$, which can be further expressed as
\begin{equation}
    \boldsymbol{y} = \boldsymbol{H} \boldsymbol{x} + \boldsymbol{\eta}
\end{equation}
where $\boldsymbol{x} \in \mathbb{C}^{M}$ denotes the feeder signals of the satellite. $\boldsymbol{\eta} \in \mathbb{C}^{K} $ denotes an additive white Gaussian noise (AWGN) vector, where each entry of $\boldsymbol{\eta}$ independently obeys the complex Gaussian distribution with zero mean and variance being $\sigma^2$, i.e., $\mathbb{E}\{\boldsymbol{\eta} \boldsymbol{\eta}^H\} = \sigma^2\boldsymbol{I}_K$. We denote $\boldsymbol{H}\in\mathbb{C}^{K\times M}$ as the downlink channel matrix between the multibeam satellite and the $K$ ground users, which can be expressed as

\begin{equation}
    \boldsymbol{H} = \boldsymbol{\Phi}\boldsymbol{C}
\end{equation}
where $\boldsymbol{C}\in\mathbb{R}^{K\times M}$ and $\boldsymbol{\Phi}\in\mathbb{C}^{K\times K}$ denote the multibeam antenna pattern and the signal phase matrix caused by different propagation paths between the satellite and the users, respectively. Since the distance between the neighbouring satellite antenna feeds is relatively small compared with the distance between the satellite and the users, it is common to assume that the phases between the user and all the feeders are the same~\cite{Christopoulos2014Multicast,Zheng2012Generic}. Therefore, $\boldsymbol{\Phi}$ is a diagonal matrix with the $k(k=1,2,\ldots,K)$th diagonal entry being $[\boldsymbol{\Phi}]_{k,k} = e^{j\phi_k}$, where $\phi_k$ denotes a random variable obeying the uniform distribution in $(0, 2\pi)$. The off-diagonal entries of $\boldsymbol{\Phi}$ are all zero, i.e., $[\boldsymbol{\Phi}]_{k,m} = 0$ for $i \neq m$. For the multibeam antenna pattern, the entry on the $k(k=1,2,\ldots,K)$th row and $m(m=1,2,\ldots,M)$th column of $\boldsymbol{C}$ can be modeled as
\begin{equation}\label{AntennaPattern}
    [\boldsymbol{C}]_{k,m} = \frac{\sqrt{G_RG_{k,m}}}{4\pi\frac{d_k}{\lambda}\sqrt{\kappa T_R B_W}}
\end{equation}
where $d_k$ denotes the the distance between the satellite and the $k$th user, $G_R$
denotes the receiving antenna gain, and $G_{k,m}$ denotes the gain between the $m$th antenna feeder and the $k$th user. $\lambda$, $B_W$, $\kappa$ and $T_R$ represent the wavelength, the bandwidth, the Boltzman constant, and the clear sky noise temperature at the receiver, respectively. Note that the function of the array fed reflector on the satellite has already been considered in the modeling of $\boldsymbol{H}$.


To mitigate the inter-beam interference caused by the satellite channel, a common preprocessing technique is introducing a precoding matrix to combat the channel distortion. Under this context, the feeder signals can be further expressed as
\begin{equation}
    \boldsymbol{x} = \boldsymbol{W}\boldsymbol{s}
\end{equation}
where $\boldsymbol{W}\triangleq [\boldsymbol{w}_1,...,\boldsymbol{w}_N]\in\mathbb{C}^{M \times N}$ denotes the precoding matrix with the $i$th column of $\boldsymbol{W}$ represented as $\boldsymbol{w}_i$, and $\boldsymbol{s} \triangleq [\boldsymbol{s}_1,...,\boldsymbol{s}_N]^T\in \mathbb{C}^{N}$ is a signal vector that contains the
symbols to be transmitted for the $N$ beams. We assume that $\boldsymbol{s}\sim\mathcal{CN}(\boldsymbol{0},\boldsymbol{I}_N)$.

For the $q(q=1,2,\ldots,Q)$th ground user in the $n(n=1,2,\ldots,N)$th satellite beam, the SINR is
\begin{equation}\label{userSINR}
    {\Gamma}_{n,q}=\frac{|\boldsymbol{h}_{n,q}\boldsymbol{w}_n|^2}{\sum_{i\neq n}^N {|\boldsymbol{h}_{n,q}\boldsymbol{w}_i|^2} +\sigma^2}
\end{equation}
where $\boldsymbol{h}_{n,q}$ denotes the $\big((n-1)Q+q\big)$th row of $\boldsymbol{H}$.

Since the power of the satellite is typically supplied by the solar wings and onboard batteries, it is significant to save the power of the satellite to extend the lifetime of onboard devices. Therefore, instead of merely improving the SE of the satellite communications, we consider improving the EE as an important metric. We define the EE as the ratio of the sum-rate of the worst ground users over the total power of the satellite. Note that we pay more attention to the worst user of each beam with respect to the fairness among different users. In a multicast system, in order to ensure the fairness of different users so that all users instead of only part of users can get satellite services, we give special concerns to the worst user of each beam.

Now we formulate the EE maximization problem of multibeam satellite communications with multicast precoding as
\begin{subequations}\label{orign_problem}
    \begin{align}
        \max_{\boldsymbol{W}}&~\frac{\sum_{n=1}^N \alpha_n \min\limits_{q=1,...,Q} \log{(1 + \Gamma_{n,q})}}{\sum_{n=1}^N \|\boldsymbol{w}_n\|^2 + P_0} \label{EEmaxObj}\\
        \text{s.t.}~&\sum_{n=1}^N \|\boldsymbol{w}_n\|^2 \leq P_T \label{TotalPowerConstraint}\\
        &~\Gamma_{n,q} \geq \overline{\Gamma}_{n},~ n=1,..,N, q=1,...,Q \label{QoSconstraint}
    \end{align}
\end{subequations}
where the numerator of the objective function in \eqref{EEmaxObj} is the weighted sum-rate of the worst ground users from each beam, while the denominator is the satellite power consumption. $\boldsymbol{\alpha} \triangleq \{\alpha_1,\ldots,\alpha_N\}$ is a set containing the predefined weights for all the $N$ beams, where the larger weight implies the higher importance of the beam, e.g., the beam covering big cities probably needs larger weight. Since the satellite power consumption is mainly determined by the platform payloads, the power consumption mainly includes that consumed by the power amplifiers for the user link, the feeder link, the remote sensing and control link, and the onboard signal processing units, which are generally denoted as $P_0$. \eqref{TotalPowerConstraint} indicates the total power constraint of the satellite communications, where $P_T$ is the maximum transmission power determined by the power amplifier on the satellite~\cite{ZLin2019Robustsecure}. Note that the power allocation is implicitly included in \eqref{TotalPowerConstraint}. \eqref{QoSconstraint} indicates the QoS constraint for each user, where $\overline{\Gamma}_{n}$ is a threshold SINR that each user in the $n$th beam should satisfy to guarantee the QoS~\cite{MAVazquez2017Non-convex}.

\section{Precoding Design}\label{sec.UserGrouping}
To solve \eqref{orign_problem}, we first introduce new variables $\gamma_{n,q}$ satisfying $\Gamma_{n,q}\geq \gamma_{n,q}\geq \overline{\Gamma}_{n} $ for $n=1,2,\ldots,N,q=1,2,\ldots,Q$. Then \eqref{userSINR} can be converted as
\begin{equation}\label{userSINR2}
    {\gamma}_{n,q}\leq\frac{|\boldsymbol{h}_{n,q}\boldsymbol{w}_n|^2}{\sum_{i\neq n}^N {|\boldsymbol{h}_{n,q}\boldsymbol{w}_i|^2} +\sigma^2},q=1,\ldots,Q,n=1,\ldots,N.
\end{equation}
We further introduce new variables $\beta_{n, q}$ satisfying
\begin{equation}\label{SINR_decompose_b}
    \beta_{n,q} \geq \sum_{i\neq n}^N |\boldsymbol{h}_{n,q}\boldsymbol{w}_i|^2 + \sigma^2.
\end{equation}
Then we have
\begin{equation}\label{SINR_decompose_a}
    \gamma_{n,q} \leq \frac{|\boldsymbol{h}_{n,q}\boldsymbol{w}_n|^2}{\beta_{n,q}}.
\end{equation}
The first-order Taylor lower bound approximation of the right side of \eqref{SINR_decompose_a} at point $\big(\boldsymbol{w}_n^{(t)}, \beta_{n,q}^{(t)}\big)$ is~\cite{qi2018precoding,ZLin2019Joint}
\begin{align}\label{SCA}
    \frac{|\boldsymbol{h}_{n,q}\boldsymbol{w}_n|^2}{\beta_{n,q}} & \geq \frac{2\text{Re}\{ (\boldsymbol{w}_n^{(t)})^H\boldsymbol{h}_{n,q}^H\boldsymbol{h}_{n,q}\boldsymbol{w}_n \}}{\beta_{n,q}^{(t)}} -(\frac{|\boldsymbol{h}_{n,q}\boldsymbol{w}_n^{(t)}|}{\beta_{n,q}^{(t)}})^2\beta_{n,q} \nonumber \\
    &\triangleq \phi^{(t)}\big(\boldsymbol{w}_n, \beta_{n,q}\big|\boldsymbol{w}_n^{(t)}, \beta_{n,q}^{(t)}\big)
\end{align}
where the right side of the inequality of \eqref{SCA} is defined as a real conditional function $\phi^{(t)}(\boldsymbol{w}_n, \beta_{n,q} | \boldsymbol{w}_n^{(t)}, \beta_{n,q}^{(t)})$. Note that we use the superscript $(t)$ to distinguish different values of variables or vectors at different iteration, since they will be iteratively updated in our algorithms, e.g., $\boldsymbol{w}_n^{(t)}$ is the value of $\boldsymbol{w}_n$ at the $t$th iteration.

Since the objective function in \eqref{EEmaxObj} is complicated due to both the maximization and minimization operations, we define $r_n \triangleq \min\limits_{q=1,...,Q}\log(1+\gamma_{n,q}),~n=1,2,\ldots,N$, which implies that
\begin{equation}\label{minRate}
  r_n \leq \log(1+\gamma_{n,q}),~n=1,2,\ldots,N.
\end{equation}

To ease the notation, we define a positive real matrix $\boldsymbol{\beta}$, where the entry on the $n$th row and $q$th column of $\boldsymbol{\beta}$ is defined as $\beta_{n,q}$, i.e., $[\boldsymbol{\beta}]_{n,q} \triangleq \beta_{n,q}, q=1,\ldots,Q,n=1,\ldots,N$. Then \eqref{orign_problem} can be converted into
\begin{subequations}\label{transform_problem}
    \begin{align}
    \max_{\boldsymbol{W},\boldsymbol{\beta}}~ &\frac{\sum_{n=1}^N \alpha_n r_{n}}{\sum_{n=1}^N \|\boldsymbol{w}_n\|^2 + P_0} \\
    \text{s.t.} ~\  &\gamma_{n,q}\geq \overline{\Gamma}_{n},n=1,...,N ,q=1,...,Q\\
    & \gamma_{n,q} \leq \phi^{(t)}\big(\boldsymbol{w}_n, \beta_{n,q}\big|\boldsymbol{w}_n^{(t)}, \beta_{n,q}^{(t)}\big) \\
    & \eqref{TotalPowerConstraint},\eqref{SINR_decompose_b},\eqref{minRate}.
    \end{align}
\end{subequations}

Note that \eqref{transform_problem} is a concave-convex fractional programming problem which can be converted into a convex problem with the Charnes-Cooper transformation~\cite{tervo2018energy,ZLin2019Robustsecrecy}. With such transformation, we define $\bar{r}_n \triangleq  \varphi r_{n}$, $\bar{\boldsymbol{w}}_n \triangleq  \varphi {\boldsymbol{w}}_n$, $\bar{\beta}_{n,q} \triangleq  \varphi \beta_{n,q}$ and $\bar{\gamma}_{n,q} \triangleq \varphi \gamma_{n,q}$, where
$\varphi$ can be regarded as an upper bound variable of the inverse of the total power consumption, i.e., $\varphi \triangleq (\sum_{n=1}^N{\left\|{\boldsymbol{w}_n}\right\|^2}+P_0)^{-1}.$ Then
\eqref{transform_problem} is converted into the following convex problem as
\begin{subequations}\label{convex_problem}
    \begin{align}
        \max_{\overline{\boldsymbol{W}},\bar{\boldsymbol{\beta}}}~ &\sum_{n=1}^N \alpha_n \bar{r}_n \\
        \text{s.t.} ~  & \sum_{n=1}^N \|\bar{\boldsymbol{w}}_n\|^2 + \varphi^2 P_0 \leq \varphi\\
        &\bar{r}_n \leq \varphi\log\left(1+\frac{\bar{\gamma}_{n,q}}{\varphi}\right)\\
        &\sum_{n=1}^N \|\bar{\boldsymbol{w}}_n\|^2\leq \varphi^2 P_T\\
        &\bar{\gamma}_{n,q} \leq \phi^{(t)}\big(\bar{\boldsymbol{w}}_n, \bar{\beta}_{n,q} \big| \boldsymbol{w}_n^{(t)}, \beta_{n,q}^{(t)}\big) \label{ConditionalFunctionConstraint}\\
        &\varphi\bar{\beta}_{n,q}\geq \sum_{i\neq n}^N |\boldsymbol{h}_{n,q}\bar{\boldsymbol{w}}_i|^2 + \varphi^2\sigma^2\\
        &\bar{\gamma}_{n,q}\geq\varphi \overline{\Gamma}_{n}
    \end{align}
\end{subequations}
which can be solved by the existing optimization tools, e.g., CVX. It is seen that given $\boldsymbol{w}_n^{(t)}, \beta_{n,q}^{(t)}$ for \eqref{ConditionalFunctionConstraint}, we can solve \eqref{convex_problem}. Suppose $\overline{\boldsymbol{W}}^*$, $\bar{\boldsymbol{\beta}}^*$, $\bar{\boldsymbol{r}}^*$ and $\varphi^*$ are the optimal solutions of \eqref{convex_problem}. We set $\bar{\boldsymbol{r}}^{(t+1)}$ the same as $\bar{\boldsymbol{r}}^*$, where $\bar{\boldsymbol{r}}^{(0)}$ is initialized to be a zero vector. Then the solutions of \eqref{transform_problem} can be obtained as $\overline{\boldsymbol{W}}^*/\varphi^*$ and $\bar{\boldsymbol{\beta}}^*/\varphi^*$, which can be used as $\boldsymbol{w}_n^{(t+1)}$ and $\beta_{n,q}^{(t+1)}$ respectively for \eqref{ConditionalFunctionConstraint} to obtain the new solutions of \eqref{convex_problem}. We iteratively run the above procedures until the stop condition is satisfied. The stop condition is set as
\begin{equation}\label{StopCondition}
  \Big|\sum_{n=1}^{N}{\alpha_n\big(\bar{r}_n^{(t)} - \bar{r}_n^{(t-1)}\big)} \Big| \leq \xi
\end{equation}
where $\xi$ is a predefined threshold to control the convergence of the energy efficiency, e.g., $\xi = 0.001$. The procedures are summarized from step~6 to step~11 of \textbf{Algorithm~1}.

\begin{algorithm}[!t]
\begin{small}
    \caption{Energy Efficient Multicast Precoding Design}
    \label{precoding_algorithm}
    \begin{algorithmic}[1]
        \STATE \emph{Input:} $\boldsymbol{H}$, $P_T$, $P_0$, $\sigma^2$, $\xi$, $\boldsymbol{\alpha}$.
        \STATE Initialize $\boldsymbol{W}_{\text{in}}$ to be any matrix satisfying (\ref{TotalPowerConstraint}).
        \STATE Initialize $\boldsymbol{\beta}_{\text{in}}$ to be any matrix satisfying (\ref{SINR_decompose_b}).
        \STATE Solve (\ref{relax_problem}) with $\boldsymbol{W}_{\text{in}}$ and $\boldsymbol{\beta}_{\text{in}}$, where the optimal values are denoted as $\boldsymbol{W}^{(0)}$ and $\boldsymbol{\beta}^{(0)}$, respectively.
        \STATE Set $t \leftarrow 0$ and $\bar{\boldsymbol{r}}^{(0)}\leftarrow \boldsymbol{0}$.
        \REPEAT
        \STATE Solve (\ref{convex_problem}) given $\boldsymbol{W}^{(t)}$ and $\boldsymbol{\beta}^{(t)}$, where the solutions are denoted as $\overline{\boldsymbol{W}}^*$, $\bar{\boldsymbol{\beta}}^*$, $\bar{\boldsymbol{r}}^*$, and $\varphi^*$.
        \STATE Set $\boldsymbol{W}^{(t+1)} \leftarrow \overline{\boldsymbol{W}}^* /\varphi^*$,
        $\boldsymbol{\beta}^{(t+1)} \leftarrow \bar{\boldsymbol{\beta}}^* /\varphi^*$.
        \STATE Set $\bar{\boldsymbol{r}}^{(t+1)} \leftarrow \bar{\boldsymbol{r}}^*$.
        \STATE Set $t \leftarrow t+1$.
        \UNTIL{\eqref{StopCondition} is satisfied}

        \STATE \emph{Output:} $\boldsymbol{W}^{(t)}$.
    \end{algorithmic}
\end{small}
\end{algorithm}

To ensure the convergence of the energy efficiency, it is important to find feasible $\boldsymbol{W}^{(0)}$ and $\boldsymbol{\beta}^{(0)}$ to start the iteration. To find feasible $\boldsymbol{W}^{(0)}$ and $\boldsymbol{\beta}^{(0)}$ satisfying the constraints of (\ref{transform_problem}), we iteratively solve the following convex optimization problem
\begin{subequations}\label{relax_problem}
    \begin{align}
        \max_{\boldsymbol{W},\boldsymbol{\beta}} ~&\sum_{n=1}^N \alpha_n r_n -\lambda\bigg(
        \psi_1 +\sum_{n=1}^N \sum_{q=1}^Q([\boldsymbol{\psi}_2]_{n,q}+[\boldsymbol{\psi}_3]_{n,q})\nonumber\\
        &+\sum_{n=1}^N([\boldsymbol{\psi}_4]_n+[\boldsymbol{\psi}_5]_n) \bigg)\\
        \text{s.t.} ~& \sum_{n=1}^N \|{\boldsymbol{w}}_n\|^2\leq P_T + \psi_1\\
        & \gamma_{n,q} \leq \phi^{(t)}\big(\boldsymbol{w}_n, \beta_{n,q}\big|\boldsymbol{w}_n^{(t)}, \beta_{n,q}^{(t)}\big)+[\boldsymbol{\psi}_2]_{n,q} \label{15c} \\
        &\sum_{i\neq n}^N |\boldsymbol{h}_{n,q}\boldsymbol{w}_i|^2 + \sigma^2\leq \beta_{n,q} +[\boldsymbol{\psi}_3]_{n,q}\\
        &\overline{\Gamma}_{n} \leq \gamma_{n,q} + [\boldsymbol{\psi}_4]_n \\
        & {r}_n\leq \log(1+\gamma_{n,q}) + [\boldsymbol{\psi}_5]_n \\
        & \psi_1\geq 0,~\boldsymbol{\psi}_2\succeq \boldsymbol{0},~\boldsymbol{\psi}_3\succeq \boldsymbol{0},~\boldsymbol{\psi}_4\succeq \boldsymbol{0}, ~\boldsymbol{\psi}_5\succeq \boldsymbol{0}
    \end{align}
\end{subequations}
where $\psi_1, \boldsymbol{\psi}_2, \boldsymbol{\psi}_3, \boldsymbol{\psi}_4, \boldsymbol{\psi}_5$ are nonnegative slack variables for each constraint and $\lambda$ is a positive penalty parameter, e.g., $\lambda=100$. We initialize $\boldsymbol{W}_{\text{in}}$ to be any matrix satisfying (\ref{TotalPowerConstraint}). We initialize $\boldsymbol{\beta}_{\text{in}}$ to be any matrix satisfying (\ref{SINR_decompose_b}). Based on $\boldsymbol{W}_{\text{in}}$ and $\boldsymbol{\beta}_{\text{in}}$, we obtain $\boldsymbol{w}_n^{(0)}$ and $\beta_{n,q}^{(0)}$ as the initialization points of \eqref{15c}. Then we iteratively solve \eqref{relax_problem} with the same procedures as iteratively solving \eqref{convex_problem}, with only difference in the stop condition, where we stop the iteration once all the slack variables are zero. Then we obtain $\boldsymbol{W}^{(0)}$ and $\boldsymbol{\beta}^{(0)}$ and set them as the input to \eqref{convex_problem} for the first iteration. The procedures are summarized from step~2 to step~4 of \textbf{Algorithm~1}.

\begin{figure}[!t]
\centering
\includegraphics[width=80mm]{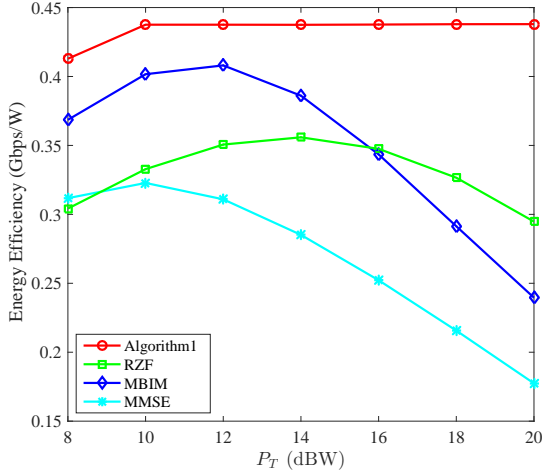}
\caption{Comparisons of the EE for different algorithms with different $P_T$.}
\label{Power}
\end{figure}

\begin{figure}[!t]
\centering
\includegraphics[width=80mm]{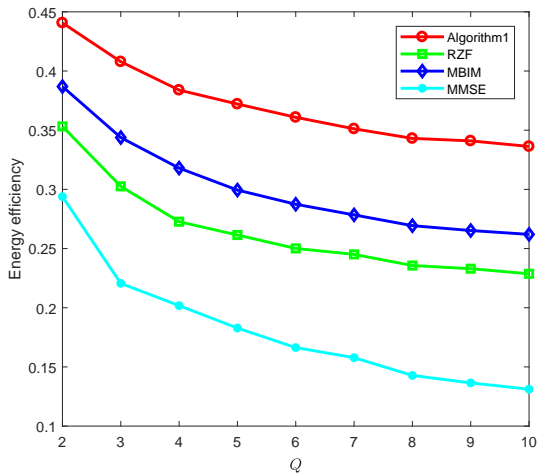}
\caption{Comparisons of the EE for different algorithms with different $Q$.}
\label{NumOfUser}
\end{figure}

\section{Numerical Results}\label{sec.NumericalResult}
To evaluate the performance of our work, we assume a multibeam satellite working in 20GHz Ka band with $M=16$ feeders and $N=16$ beams. Within the coverage of the satellite, all users are uniformly distributed. We set $\alpha_1=\alpha_2=\cdots=\alpha_N=1$. The channel used in our simulation is provided by the European Space Agency (ESA).  Since we normalize the noise power by $\kappa T_R B_W$ in \eqref{AntennaPattern}, we set $\sigma^2=1$~\cite{Joroughi2017Generalized}. The detailed parameters are listed in Table \ref{parameters_table}. The Boltzmann constant is $\kappa=1.38 \times 10^{-23}$~J/K. In the following, we compare \textbf{Algorithm~1} with the multibeam interference mitigation (MBIM) algorithm~\cite{Joroughi2017Generalized}, regularized zero-forcing (RZF) algorithm~\cite{stankovic2008generalized}, and minimum mean squared error (MMSE) algorithm~\cite{cottatellucci2006interference}.

\begin{table}[h]
\centering
\caption{Simulation Parameters}\label{parameters_table}
\begin{tabular}{c|c}
\Xhline{.8pt}
Parameter & Value \\ \hline
Satellite height &  35786km\\
Carrier frequency & 20 GHz (Ka band) \\
Total bandwidth ($B_W$) & 500 MHz  \\
User antenna gain & 41.7 dBi \\
$G/T$ & 17.68 dB/K \\
Feed Radiation pattern & Provided by ESA \\
\Xhline{.8pt}
\end{tabular}
\end{table}
As show in Fig. \ref{Power}, we compare the EE for different algorithms with different $P_T$. We set $Q =2 $ and $P_0 = 10 \text{dBW}$. For the RZF, MBIM and MMSE algorithms, the EE curves increase at first and then decrease. The reason is that the increment of power consumption is faster than that of sum-rate. On the contrary, the curve of \textbf{Algorithm~1} can increase without any drop. It is shown that the EE cannot be always improved by merely increasing the power of the satellite. When $P_T=10\text{dBW}$, \textbf{Algorithm~1} can achieve almost the maximum of the EE. Since the optimized feasible points are initialized for \textbf{Algorithm~1}, it can achieve fast convergence within a smaller number of iterations. As show in Fig. \ref{NumOfUser}, we compare the EE for different algorithms with different $Q$. We set $P_T=14 \text{dBW}$ and $P_0 = 10 \text{dBW}$. It is seen that the EE curves decrease as $Q$ gets larger. With more users in each beam, the channel coherence among different users increases, and consequently the SINR of the worst ground user in each beam will become smaller, which results in the lower sum-rate and thus the smaller EE.

\section{Conclusions}\label{sec.conclusion}
We have presented an iterative algorithm to design the energy efficient precoding for multiuser multibeam satellite communications. The algorithm has been verified by the measured channel data of multibeam satellite communications. The future work will be continued with the focus on finding a proper balance between the number of served users and the sum-rate.

\bibliographystyle{IEEEtran}
\bibliography{IEEEabrv,IEEEexample}

\end{document}